# Evaluating Power-to-Heat-to-Power Storage Potential in Wind- and Solar-Dominated Energy Markets

Alicia López-Ceballos[1*], Alejandro Datas[1*], Ignacio Antón[1], Carlos Cañizo[1] and Marta Victoria[2]

[1] Instituto de Energía Solar, Universidad Politécnica de Madrid, Avenida Complutense, 30, 28040, Madrid, Spain

[2] Department of Wind and Energy Systems, Technical University of Denmark, Elektrovej, 325, 2800, Lyngby, Denmark

* Corresponding author 1: a.l.ceballos@upm.es

* Corresponding author 2: a.datas@upm.es

## Abstract

This study assesses the role of Power-to-Heat-to-Power Storage (PHPS) systems, also known as Carnot batteries, in the national energy system of a wind-dominated (Denmark) and a solar-dominated (Spain) region. Using the open-source PyPSA framework, we model sector-coupled electricity and heating systems, and evaluate individual and district heating configurations for heat provision at residential level, with and without PHPS waste heat recovery. Our results show that PHPS is most viable in individual heating systems, and in wind-dominated locations, where its low cost per energy capacity enables it to balance long-duration fluctuations. Moreover, waste heat recovery is essential to enhance PHPS competitiveness, allowing it to displace lithium-ion batteries. Conversely, PHPS is largely outcompeted in district heating systems, which favour low-cost centralized energy storage. Sensitivity analysis highlights that PHPS viability depends primarily on achieving low energy capacity costs (≤10 €/kWh) and maintaining reasonable heat-to-power conversion efficiencies. Overall, while PHPS complements existing storage technologies by providing dispatchable electricity and heat, its potential adoption is highly context-dependent, influenced by climate, heating system configuration, and competing storage costs.

# 1. Introduction

The global capacity of Variable Renewable Energy (VRE), including solar and wind, is 3 TW in 2024 [1] and is projected to reach 30 TW within the next 25 years [2], at which point it would represent more than 80% of total world electricity generation capacity. However, the inherent intermittency of these sources and the temporal mismatch between generation and demand necessitate effective balancing measures. Sector coupling, particularly through heat electrification, has been identified as a cost-effective pathway for integrating VRE and decarbonize the energy sector [3], [4]. Yet, while the additional power loads induced by heat electrification can provide flexibility that initially reduces storage requirements, energy storage systems remain essential to cost-effectively balance fluctuations in solar and wind generation as renewable penetration increases [2].

Although Pumped Hydro Storage (PHS) currently dominate global energy storage installed capacity [5], [6] falling costs of lithium-ion (Li-ion) batteries have accelerated their deployment, accounting for 80 % of newly installed battery capacity in 2023 [6]. The techno-economic characteristics of Li-ion batteries – characterized by a high cost per energy (CPE) capacity, a relatively low cost per power (CPP) capacity and a very high round-trip efficiency – make them particularly well suited for addressing short-duration or intra-day fluctuations [4], [7], [8], especially in solar-dominated regions. Nevertheless, achieving a high share of VRE cost-effectively will require complementing short-duration solutions with long-duration energy storage (LDES) [7]. LDES technologies are distinguished by their low CPE, making them crucial for addressing extended periods of low renewable generation.

A notable example of LDES technology is Power-to-Heat-to-Power Storage (PHPS), also known as Carnot battery [9] or Electro Thermal Energy Storage (ETES) [10], among others. In PHPS systems, electricity is converted into heat via a Power-to-Heat (P2H) converter and stored as either sensible [11] or latent [12] heat at high temperatures. The stored heat is then converted back into electricity using a Heat-to-Power (H2P) converter.

This study is focused on PHPS systems that use electrical resistance heating to convert electricity into heat [13], [14], [15], and therefore does not consider the potential use of heat pumps as H2P unit [16]. For thermal storage, low-cost materials such as ferrosilicon alloys [15], graphite blocks [17], [18], other ceramics ([19]), clay bricks [20], or stones ([21]), have been used, insulated from the environment with inexpensive porous refractory materials. The combination of low material cost (in €/kg) and high energy density (in Wh/kg) yields a low CPE (in €/Wh), a key requirement for LDES applications [15]. The H2P converter is a critical component of the system, as it largely determines the CPP and contributes most significantly to the round-trip conversion efficiency. Energy losses from this conversion, released as low-temperature heat, can be recovered; and similar to a Combined Heat and Power (CHP) plant, PHPS can deliver both heat and electricity in their output.

Companies like 1414 Degrees [22], E2S Power [18], Rondo [23] or Gamesa [21] have suggested a storage technology combined with well-established turbogenerators (e.g., steam turbines) as H2P converter. Alternatively, other companies like Antora [14], the Fourth Power [24], or Thermophoton [25], are developing solid-state alternatives based on Thermophotovoltaics (TPV) [26], [27].

Previous techno-economic analyses of PHPS systems, charged via resistive heating, have primarily focused on their feasibility when included in residential settings. These studies often examine applications such as reducing natural gas consumption [28], [29] or their role in fully electrified buildings [15], [30], frequently comparing PHPS with Li-ion batteries. Peng et al. [11] studied the use of a graphite based PHPS system with TPV for energy discharge in the steel and iron industry, highlighting its potential for long duration energy storage. Other research has looked into hybridizing PHPS with Li-ion batteries to boost the synergies of both storage technologies [30]. Other works have also analysed the impact of leveraging waste heat from H2P conversion in large building applications [31]. On a broader scale, several works have assessed the feasibility of PHPS systems storing energy in molten salts, comparing their performance to other configurations including Li-ion batteries, wind turbines or PV installations [10], [32]. Most of the aforementioned works rely on techno-economic models that optimize the system sizing, with the exception of those by Okazaki [10] and Peng et al. [11].

Prior assessments of PHPS economic feasibility assumed that its integration has a negligible impact on electricity prices and the overall energy system configuration, thereby limiting their relevance for system-level analysis. In contrast, this study integrates PHPS into a stylized, country-wise energy system model in Europe. Our model includes wind- and solar-dominated systems, sector coupling with heating, and evaluates the impact of waste heat utilization. Within this comprehensive framework, this analysis addresses the following research questions:

- What is the optimal capacity and operation of PHPS under different scenarios?
- How do varying technical and economic assumptions influence these outcomes?

For the low-temperature heating sector (i.e., domestic hot water), we assume complete electrification through heat pumps and heat resistors. PHPS is analysed in combination with two types of low-temperature thermal storage systems: centralized thermal energy storage (CTES), often installed in district heating systems in Baltic and Scandinavian countries, and to a lesser extent, some Eastern and South-Eastern European regions [35], [36], [37]; and individual thermal energy storage (ITES), which is more typical in decentralized heating systems in Southern

European countries [33]. Both configurations are evaluated in two representative contexts: wind-dominated systems, characteristic of Northern Europe, and solar-dominated systems, characteristic of Southern Europe.

This article is divided into 7 main sections: after the introduction in section 1, the methodology and the modelled scenarios are described in Section 2. Results and Sensitivity Analysis are summarized in Sections 3 and 4, while Discussion is conducted in Section 5. The limitations of the study are included in Section 6. Finally, conclusions are gathered in Section 7.

## 2. Methodology

The model developed using the open-source software PyPSA [34], simulates the operation of a single European country energy system of a full year with hourly resolution under perfect foresight. The system incorporates both power and heating sectors, each composed of various elements such as generators, demand loads and energy storage units. The capacities of these components (e.g., generators or energy storages) and their dispatch are jointly optimized to meet the demands, while minimizing the annualized total system cost. A key constraint in the optimization is the limitation of $CO_2$ emissions to 5% of the 1990 values for each country. The model's behaviour is governed by a set of equations and constraints, which are detailed in Appendix A. Additional information on PyPSA modelling tool, and its applications can be found in studies, such as [4], [35], [36], [37].

In this study we have compared two locations with different climate conditions in Europe. The wind-dominated location, represented by Denmark, is characterized by having high wind-generation and high heating demand. Conversely, the solar-dominated location, exemplified by Spain, has higher solar generation, and reduced heating demand. The data used for the simulations is plotted in Figure 1 and Figure B. 1 for Denmark and Spain, respectively. Panels a and b in Figure 1 depicts the hourly capacity factor (CF) data of solar and wind generation, respectively, retrieved from [38], [39]. Panels c and d show the electricity and heating demand of the country normalized to their respective average yearly demand from [40]. Wind generation in both climates exhibits synoptic fluctuations, typically spanning a couple of days, with higher output during the winter months, correlating well with heating demand (panel d). As expected, solar generation in both locations demonstrates daily variations. Each country is modelled as an isolated copperplate node, meaning that energy transmission between countries (exports or imports) and bottlenecks within the country itself are neglected.

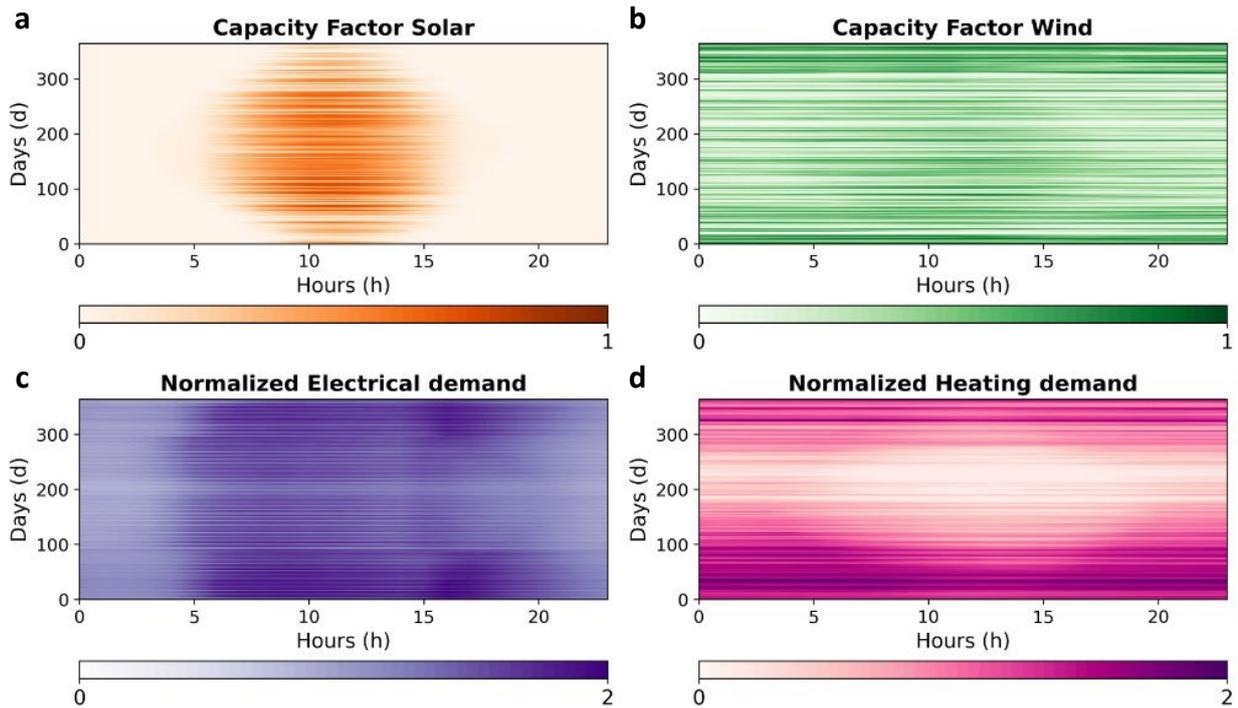

Figure 1. Panels a-d plot hourly data throughout the day (x-axis) and the year (y-axis) for the wind-dominated location, Denmark. Panel a and b represent the solar and onshore wind capacity factors respectively. Panels c and d contain the normalized power and heating demand relative to their respective average demand.

Figure 2 depicts the modelled energy system, comprised of generators, energy storage systems and demands connected to two buses: electricity and heat. The heating sector is assumed to be fully electrified, and both buses are connected via heat pumps and heat resistances, which convert electricity into heat with certain efficiency. The power generators, storage units and demand are attached to the electrical power bus. The wind and solar generators are defined by a CF, which is a value between 0 and 1, representing the energy that can be potentially generated over an hour by each generator normalized to its nominal power capacity. An Open Cycle Gas Turbine (OCGT) generator is included to represent other existing backup technologies simplifying the energy system. It is worth noting that, if the $CO_2$ emissions were not constrained to 5% of the 1990 values for each country, OCGT would become the predominant technology to balance variable renewable energy fluctuations displacing energy storage systems.

There are two types of electrical storage systems: Li-ion batteries, and PHPS. In addition to charging and discharging into the electrical power grid, the waste heat generated at the H2P converter during the PHPS discharge is directed to the heating bus. In different scenarios, one of the two types of low-temperature energy storage systems is considered: CTES and ITES. The former represents a centralized thermal energy storage solution, typically used in district heating systems. The latter is modelled as an individual water tank, with considerable standing losses (33 %/day meaning it is fully self-discharged within 3 days), which will ultimately limit its capacity to act as seasonal storage. Both heat stores (CTES and ITES) can be charged using either the resistive heater, the heat pump, or the waste heat from the H2P converter within the PHPS system.

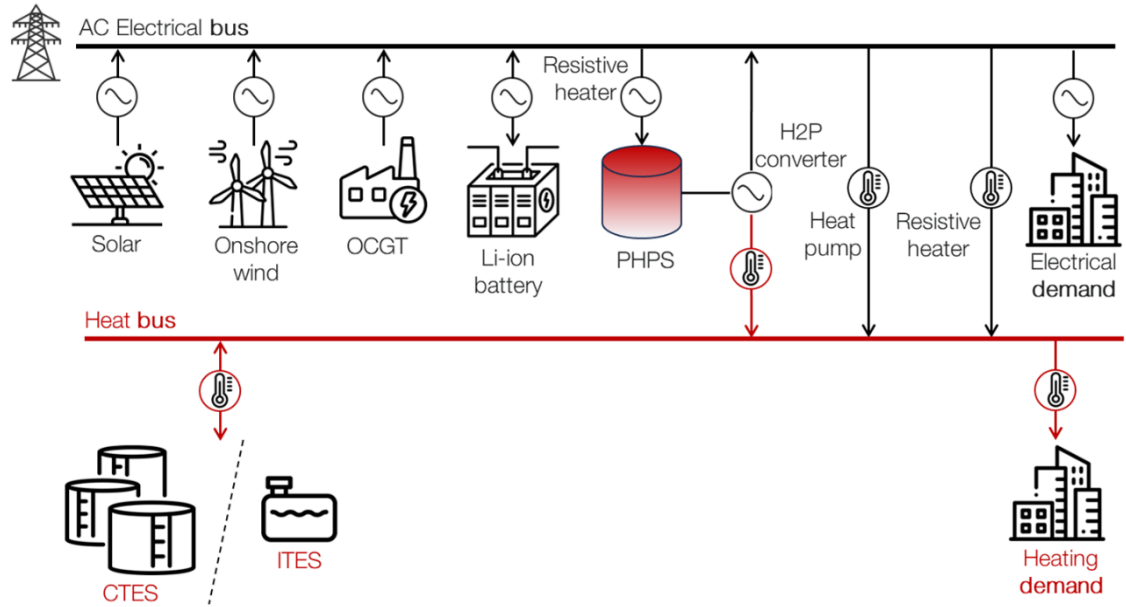

Figure 2. Overview of the modelled energy system including the electrical bus with the generators and storages, and the heating bus with the thermal energy storages.

In addition to analysing the influence of the region on PHPS feasibility, we consider three configurations: (1) A base case, with neither CTES or ITES; (2) decentralized, representing PHPS coupled with individual heating systems and including only ITES tanks; and (3) centralized, representing PHPS integrated into a district heating system and including only the CTES tanks. Each configuration is evaluated under two use-case scenarios: "waste heat," which includes waste heat utilization, and "no waste heat," which excludes it, yielding a total of 12 combinations.

A distinctive feature of PHPS is its simultaneous production of electricity and heat, with their ratio fixed by the H2P conversion efficiency, assumed constant at 0.3 in all simulations (Table 1). In the "waste heat" scenarios, this coupling means that PHPS discharge is constrained by the lower of the two demands: if no heat sink is available, either through immediate use or storage in CTES or ITES units, the system cannot be discharged, even when electricity demand exists.

To facilitate the comparison between scenarios, we use the number of average hourly load (av.h.l) that can be covered with the optimized PHPS capacity. This value represents the hours of average power demand that could be covered by discharging the energy storage capacity. As a reference, the currently existing pumped hydro storage capacity in Spain is approximately 70 GWh [41], which considering and average power demand of 28 GW translates into to 2.5 av.h.l. When the optimal energy capacity for PHPS is below 2 av.h.l, PHPS is considered not contributing significantly to the energy system. It should be noted that this metric emphasizes the role of storage technologies with relatively high energy capacity. Consequently, high-power but low-energy technologies, characterized by low av.h.l values yet potentially significant for mitigating short-term fluctuations, are not fully captured by this indicator. Nonetheless, this approach is appropriate for the present study, which focuses on system-level adequacy and renewable integration rather than power quality.

The technical and economic assumptions for each component, corresponding to the year 2030, are detailed in Table 1. The assumed marginal cost for OCGT is 0.025 €/kWh$_{th}$, determined by fuel cost. A marginal cost of 0.01 €/MWh is applied to the discharging of Li-ion battery and PHPS, and to the heat resistance to prevent unusual operation due to excess of renewable generation. The financial discount rate to annualize components cost is 7 %. The cost of the P2H converter of the PHPS is estimated as the sum of the costs of CTES charging power [42], and the P2H converter for high temperatures [15], [43]. Unknown economic parameters related to PHPS, such as the Fixed Operation and Maintenance (FOM) costs or lifetime, have been taken from molten salts [42].

Table 1. Technical and economic assumptions for the components included in Figure 2.

|  | Cost | Unit | FOM (%/y) | Lifetime (y) | Efficiency (%) | Standing loss (%/day) | Ref. |
|---|---|---|---|---|---|---|---|
| Generators |  |  |  |  |  |  |  |
| Solar PV | 543 | €/kW$_{el}$ | 2 | 40 |  |  | [42] |
| Wind onshore | 1100 | €/kW$_{el}$ | 1.2 | 30 |  |  | [42] |
| OCGT | 450 | €/kW$_{el}$ | 1.8 | 25 | 42 |  | [42] |
| Energy storage system |  |  |  |  |  |  |  |
| AC-DC converter Li-ion | 169 | €/kW$_{el}$ | 0.3 | 10 | 95 |  | [42] |
| Energy storage Li-ion | 150 | €/kWh | 0 | 25 |  | 0 | [42] |
| H2P converter PHPS | 1000 | €/kW$_{el}$ | 0.27 | 35 | 30 |  | [15] |
| P2H converter PHPS | 88 | €/kW$_{el}$ | 1.7 | 35 | 100 |  | [15], [42] |
| Energy storage PHPS | 10 | €/kWh | 0.3 | 35 |  | 5 | [15] |
| CTES tank | 0.6 | €/kWh | 0.6 | 25 |  | 0.6 | [42] |
| ITES tank | 20 | €/kWh | 1 | 20 |  | 33 | [42] |
| Others |  |  |  |  |  |  |  |
| Heat pump | 2700 | €/kW$_{el}$ | 0.2 | 25 | 360 |  | [42] |
| Heat resistance | 68 | €/kW$_{el}$ | 1.7 | 20 | 100 |  | [42] |

# 3. Results

Figure 3 contains the optimal Li-ion, PHPS and CTES or ITES energy storage capacity for the different configurations as a percentage of the total energy capacity within the optimal system. We compared results for energy systems of wind (panels a and b) and solar dominated regions (panels c and d) including (panels b and d) or not (panels a and c) the possibility of using the waste heat from the H2P conversion from PHPS. Optimal energy capacity, expressed in av.h.l, for the PHPS is included in Table 2. Optimal energy capacities of rest of the energy storage systems (Li-ion battery, CTES and ITES) are shown in Table B. 1.

## 3.1. PHPS selected for individual heating systems

We begin with the decentralized configuration, where PHPS proves to be a viable energy storage solution, regardless of the location and the use of waste heat. Its role is particularly pronounced in wind-dominated regions, as reflected in the higher investment in PHPS energy capacity shown in Table 2. This outcome is driven by the relatively low cost per unit of energy capacity, which makes PHPS well suited for long-duration storage that aligns with the low-frequency fluctuations characteristic of wind generation.

In wind-dominated regions, where heat demand is higher, the availability of ITES slightly reduces the optimal PHPS energy capacity, as ITES acts as a buffer between heat production and demand. This allows a greater share of surplus electricity to be converted directly into heat via heat pumps, a more efficient form of thermal energy storage. In solar-dominated regions, however, ITES has the opposite effect, slightly increasing the optimal PHPS capacity. Here, PHPS was previously constrained by the absence of a heat sink, which limited discharge. With ITES available, waste heat can be stored, enabling greater utilization of PHPS and thus a larger optimal energy capacity.

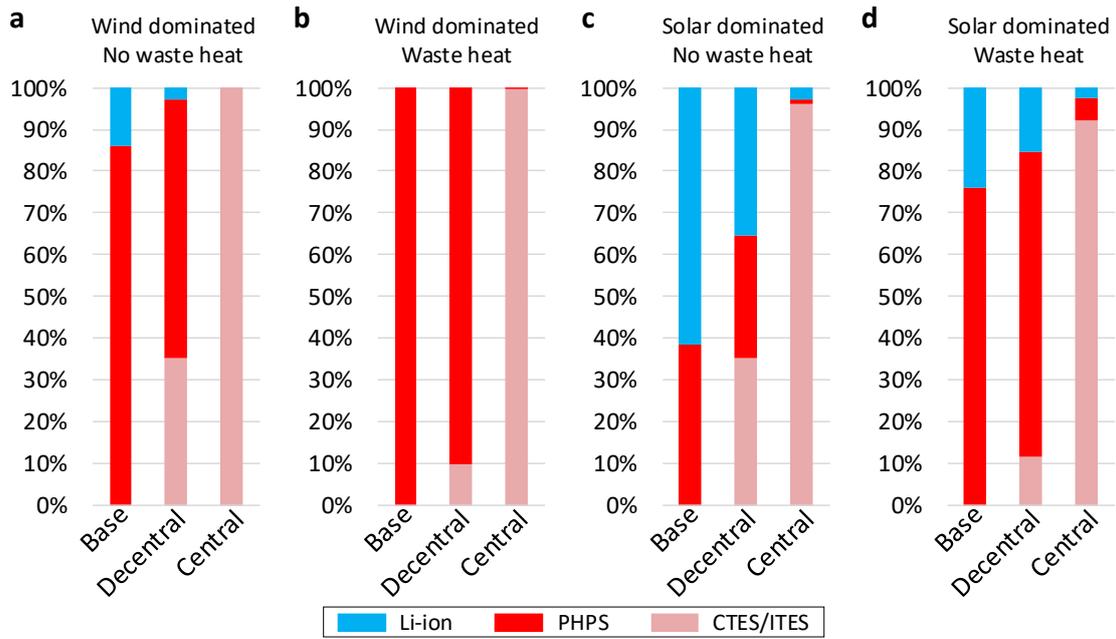

Figure 3. Optimal energy capacities of the Li-ion (blue) battery, PHPS system (red) and CTES or ITES (pink) as a percentage of the whole energy capacity within the system. Panel a and b depict results for the wind dominated region and panel c and d for the solar dominated region, when waste heat is not used (panels a and c) and when it is used (panels b and d). Each panel depict the results for the Base, Decentralized and Centralized configurations.

Table 2. Optimal PHPS energy capacity quantified as average hourly load (av.h.l) that can be covered when fully discharged. This parameter is used as an indication of the relevance of the PHPS storage.

| Configuration \ Location | Wind-dominated | | Solar-dominated | |
|---|---|---|---|---|
| | No waste heat | Waste heat | No waste heat | Waste heat |
| Base | 25 | 49 | 5 | 16 |
| Decentralized | 23 | 46 | 5 | 19 |
| Centralized | 0 | 2 | 2 | 8 |

To further illustrate the dynamic performance of the system, panels a and b in Figure 4 depict the electricity demand normalized by the yearly average electricity demand in the wind-dominated location, Denmark (3.7 GW) for the ITES configuration without using waste heat. These panels show the generation by each energy technology (positive values) and the electricity used to charge the electrical stores (negative values) for a three-day period in winter (panel a) and summer (panel b). Notice that wind generation for this location is higher in winter than in summer, while solar generation shows the opposite trend (panels a and b in Figure 1).

During winter days (panel a), long-time fluctuations of wind generation (>20 h) are balanced by PHPS, which is charged and discharged over periods of 20-40 h. Remarkably, PHPS is charged at higher power than it is discharged, making it capable of storing large amounts of energy for extended periods while releasing it more gradually. The possibility of decoupling charging and discharging power capacities, allows the system to invest in higher charging power and energy capacities, taking advantage of their relatively low cost, while reducing the discharging power, given its higher cost. As a result, the optimal PHPS configuration has a discharge duration at maximum discharging power capacity of ~42 h (panel d), well suited to balance wind fluctuations. This is reflected in the synoptic behaviour of its normalized State of Charge (SOC) shown in panel c (Figure 4) and in the charging and discharging patterns during the winter months illustrated in panel d. PHPS is also used to cover base electricity demand, seen by the constant and low PHPS discharge power in panels a and b.

Notably, panels b and d show that PHPS also balances the daily solar generation during the summer months. Therefore, PHPS can also be used for shifting solar peak generation at midday, Although Li-ion batteries prevails in solar-dominated locations (see Figure B. 2 and Table B. 1), PHPS is also present to manage daily variations. Therefore, PHPS is used simultaneously to balance long-duration wind fluctuations and daily fluctuations from solar generation.

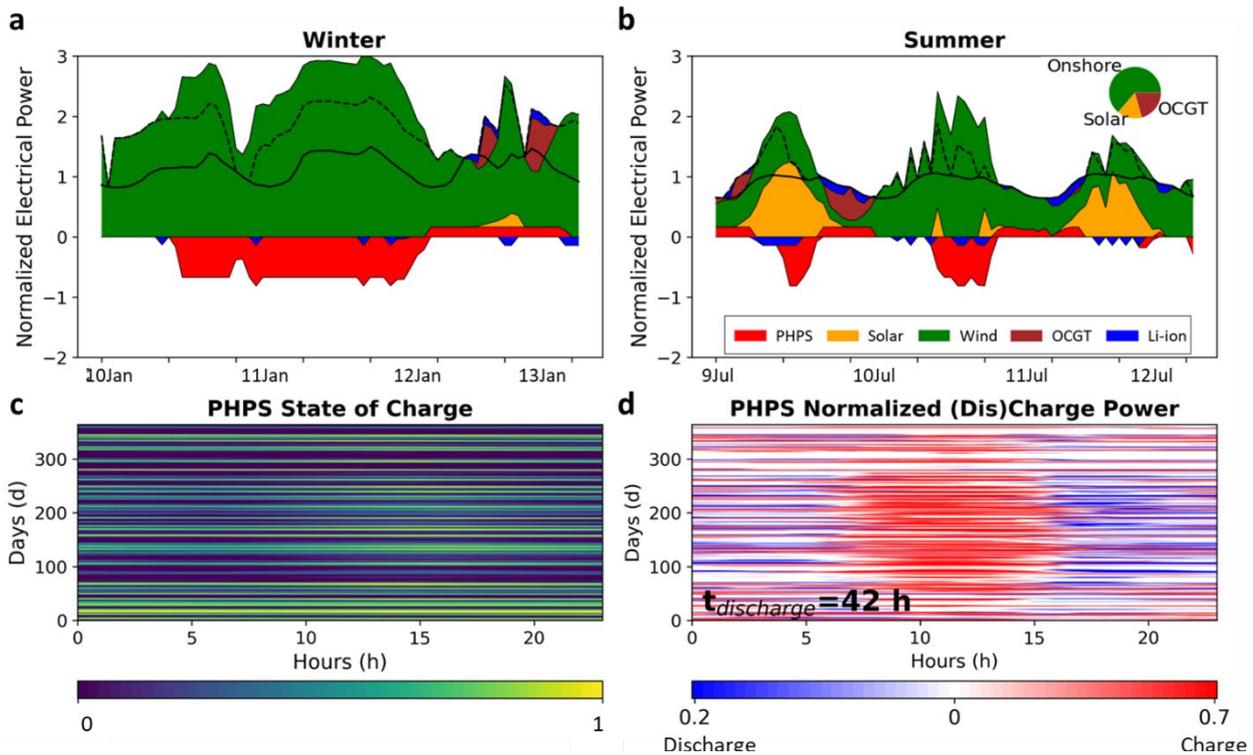

Figure 4. Energy balance and storage operation for the wind-dominated location, not considering waste heat in the decentralized configuration. All the panels in this figure show values normalized to the average electrical demand for Denmark (3.7 GW). Panels a and b depict the normalized electrical generation (positive values) of wind (green), solar PV (yellow), OCGT (brown), PHPS (red) and Li-ion batteries (blue) for three days in winter (panel a) and summer (panel b). Negative values represent the power used to charge the PHPS (red) and Li-ion batteries (blue). The normalized exogenous electricity demand is plotted in a thick black line. The normalized total

electricity, represented with dashed black line, includes electricity demand and heat pump and heat resistance electrical consumption. The annual electricity mix for this scenario is plotted in a pie-chart, inserted in panel b. Panel c shows the normalized state of charge of the PHPS throughout the day (x-axis) and year (y-axis), being yellow when completely charged and dark-blue when discharged ($E_{cap-PHPS}$=85.1 GWh). Panel d shows the normalized charging (red) and discharging (blue) power of the PHPS throughout the day (x-axis) and year (y-axis). Its discharge time at maximum discharging power capacity, $E_{capac-PHPS}$ (kWh)·$\eta_{disch-PHPS}$/$P_{el-disch-PHPS}$ (kW), is 42 hours.

### 3.2. Using waste heat increases the value of PHPS

When considering the use of the waste heat (panels b and d in Figure 3), PHPS optimal energy capacity is boosted in all the configurations, as shown in Table 2. Moreover, the appearance of PHPS also results in a significant reduction of the required Li-ion battery capacity (Figure 3), especially in wind-dominated regions where Li-ion battery is absent in all scenarios and fully substituted by PHPS (panel b in Figure 3). This is associated with the presence of wind generation, better suited with longer duration energy storage systems, and a better use of the waste heat due to the higher heating demand than in solar-dominated regions.

Remarkably, when waste heat from PHPS is leveraged, simultaneous charging and discharging of PHPS occur in both solar and wind-dominated systems (e.g., solar-dominated, panel b in Figure B. 3). This represents a net production of heat (i.e., waste heat) from electricity, which coincides with periods of peak heating demand (i.e., the coldest winter months). During these peaks, the heat resistance and the heat pump operating at their maximum power capacities are insufficient to meet the heating demand, and the waste heat from PHPS provides the additional supply needed. This operation is exemplified in Figure B. 3, which depicts some days at the beginning of the year for the solar-dominated scenario.

### 3.3. PHPS and Li-ion displacement through the integration of CTES

In scenarios where CTES is available, PHPS is barely selected, with the exception of PHPS systems that leverage waste heat in solar-dominated regions. The lack of PHPS in the rest of scenarios is attributed to the low energy capacity cost of CTES, which allows it to store large amounts of heat, efficiently generated from heat pumps, at a competitive price. This economic advantage enables the installation of large thermal energy storage with low stand-by energy losses, making it effective as a LDES solution and hindering the selection of PHPS. The advantage of CTES is especially remarkable in wind-dominated regions, where neither Li-ion batteries nor PHPS (Figure 3 panels a and b) are present in the optimal solution, and CTES is the sole selected storage technology. This outcome stems from the temporal correlation between heating demand and wind availability, combined with the fact that PHPS is not cost-competitive as a long-duration storage option in these regions. Therefore, low-cost CTES provides an effective means of managing these low-frequency fluctuations. Surprisingly, no electrical storage is included in this scenario, underscoring that in wind-dominated regions, heating demand is a more critical driver, as this demand is met instantaneously by direct wind and solar generation.

The only situation where PHPS significantly coexist with CTES is in solar-dominated regions and when the waste heat from PHPS is leveraged. As discussed in the previous sub-section, this is attributed to the ability of generating both heat and electricity in contrast with the CTES. However, even in this case, there is one order of magnitude difference between CTES and PHPS optimal energy capacities, (refer to Table B. 1 in Appendix B), indicating the key role of low-cost CTES to absorb both solar and wind surplus generation. The main contribution of PHPS in this case is to enable a reduction of both Li-ion and CTES capacities (Table B. 1), providing the system with a storage capable of supplying electricity for long duration applications.

## 4. Sensitivity analysis

In this section, we aim to assess the robustness of our results and determine whether any uncertain techno-economic assumptions are hindering PHPS feasibility. For the sake of simplicity, we focus this analysis on the wind-dominated system comparing decentralized (represented by the square marker in the following figures) and centralized (circle marker) configurations, when not using the waste heat (blue line) or when using it (red line). Figure 5 below presents a sensitivity analysis of six different parameters: (a) cost per energy capacity $CPE_{PHPS}$, (b) standing losses, (c) cost per discharging power capacity $CPP_{PHPS-disch}$, (d) cost per charging power capacity $CPP_{PHPS-ch}$, (e) H2P converter efficiency and (f) Li-ion battery cost assumption. Results for solar-dominated system, included in Figure B. 4 in Appendix 12, follow similar trends to wind-dominated locations.

As a prominent result, the PHPS cost per energy capacity $CPE_{PHPS}$ in panel a is the most sensitive parameter, being imperative for PHPS to have CPE≤10 €/kWh to enhance its feasibility. These energy capacity costs align with results found in literature [7], [36], [44], [45], which state that LDES requires CPE<20 €/kWh to be feasible. Remarkably, when PHPS uses the waste heat with CTES (circle red line), it needs $CPE_{PHPS}$ ≤~3 €/kWh to be selected. Thereby, in addition to the importance of using waste heat from PHPS, significantly reducing the energy cost is also imperative for PHPS to be feasible in wind-dominated systems with CTES. These results align with Sepúlveda et al. [44], who claim that when competitive technologies (gas with carbon capture storage, or blue hydrogen) are present in the energy system, CPE needs to be very low (1 €/kWh).

One of the possible strategies to achieve low PHPS energy costs is reducing the insulation investment [15]. Panel b in the same figure illustrates that slightly increasing the self-discharge losses up to 10 %/day, thereby potentially reducing CPE, does not have an impact as significant on PHPS' economic viability, agreeing with recommendations by Peng et al. [11]. Moreover, given the thermal properties of PHPS, increasing the scale of the energy capacity reduces its cost [15]. PHPS developers are encouraged to make a thorough analysis on the benefit of reducing CPE at the expense of increasing thermal stand-by losses depending on the size of the energy storage. Following a trend similar to standing-by losses, the charging power cost in panel c shows little change while $CPP_{ch}<\sim100$ €/kW.

By having low energy and charging power costs, PHPS can tolerate high $CPP_{disch}$ (panel d) with not very high discharge efficiency (panel e). This results into low discharge power capacity thereby, PHPS is operated at full capacity for long periods. Despite reducing $CPP_{disch}$ (<300 €/$kW_{el}$) does not increase the presence of PHPS, panel d evidences that its deployment is significantly hindered when going above 1000 €/$kW_{el}$. Conventional H2P converters like Brayton or Rankine cycles can attain these costs (<1000 €/$kW_{el}$) in large-scale devices [46], but not in small-scale applications. Developers of solid-state based PHPS, such as thermophotovoltaic (TPV) cells (e.g., LHTPV battery [15] or Thermal Grid Energy Storage [17]), predict a potential discharging cost in the range of 200-1000 €/$kW_{el}$ regardless of the scale.

Despite the PHPS system can tolerate high discharging costs, its feasibility is severely hindered at discharging efficiencies below 30 %, as seen in panel e. This is true except for the decentralized configuration using waste heat, in which the PHPS feasibility barely depends on reducing the efficiency, allowing up to a 3 % of discharging efficiency even in solar-dominated systems (see panel e in Figure 5 and Figure B. 4). This is because the large amount of waste heat is used to instantaneously cover the heating demand. As a reference, Brayton and Rankine cycles operate at 25-40% efficiency [46], while current experimental TPV devices surpass 40% according to literature [47].

As a summary of the sensitivity analysis, we learned that PHPS developers should focus on reducing energy capacity cost and limiting H2P efficiency drops. On top of that, reductions in Li-ion's future cost can significantly hinder its feasibility, agreeing with results found by Liu et al. [32]. Panel f shows the PHPS is not selected at all when waste heat is not utilized with only a 40% reduction with respect to Li-ion battery cost projected for 2030 [42].

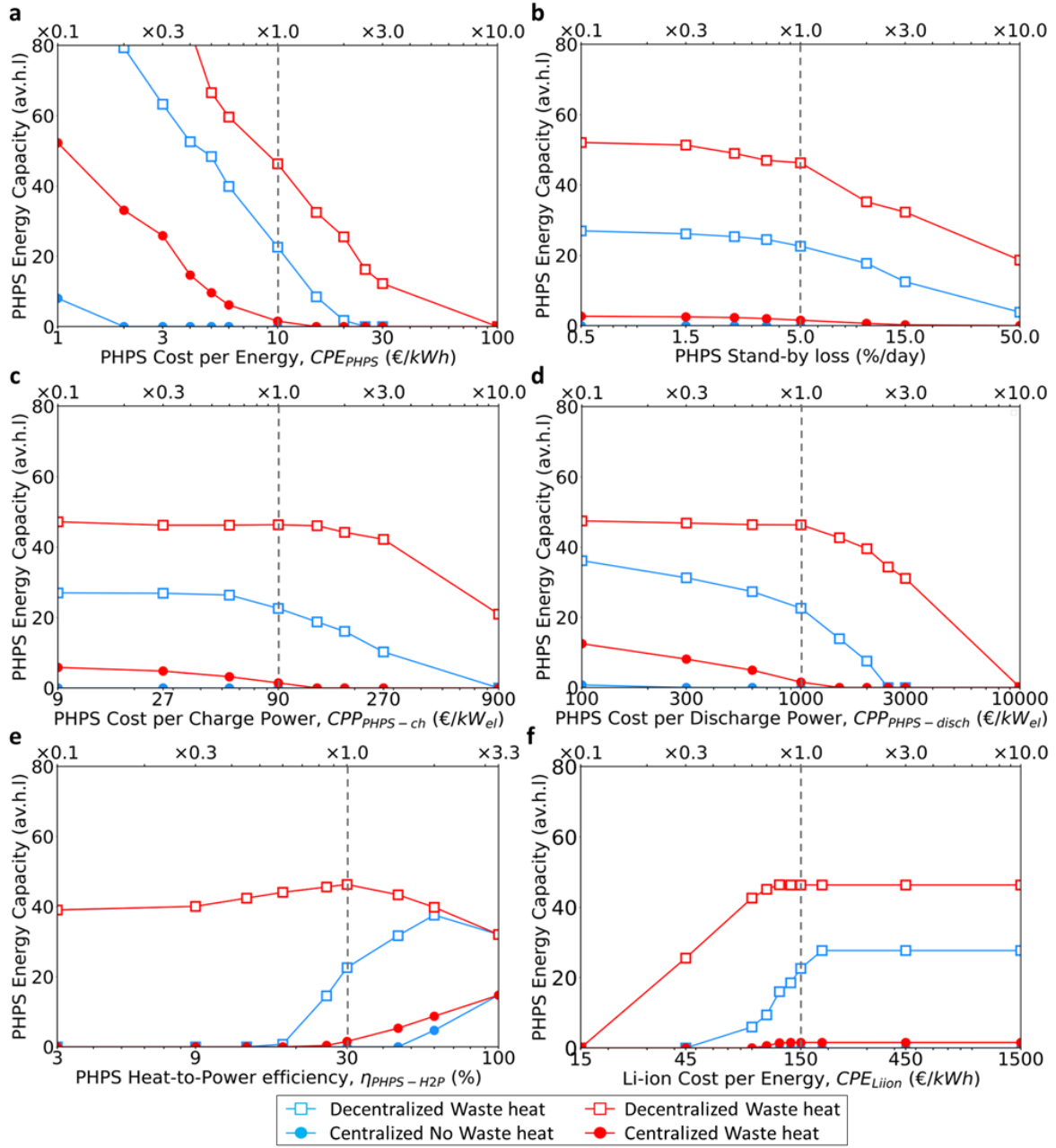

Figure 5. Optimal PHPS energy capacity for wind-dominated system, expressed in av.h.l, as a function of deviations from the reference values of different parameters: (a) Cost per Energy (CPE, €/kWh), (b) stand-by losses (%/day), Cost per (c) Discharge and (d) Charge electrical Power ($CPP_{dis}$, $CPP_{ch}$, €/kW), (e) Heat-to-Power conversion efficiency (H2P, %), and (f) Li-ion cost variation. The reference techno-economic assumptions in Table 1 are depicted by the dashed grey line.

## 5. Discussion

Our analysis shows that PHPS deployment is especially suitable when CTES is unavailable. In such cases, PHPS is selected in all scenarios, regardless of region or waste heat utilization. Its applicability is particularly relevant in wind-dominated systems and when PHPS waste heat is recovered. However, wind-dominated regions typically correspond to northern countries (Baltic and Scandinavian) with established district heating systems [33], which makes the centralized configuration a more favourable option for balancing the system, and, consequently, leads to PHPS not being selected.

In a previous study, Victoria et al. [4] concluded that optimal solutions in Europe include both CTES and hydrogen to balance long-duration fluctuations of wind generation. This finding differs from our outcomes, where PHPS is not part of the optimal solution in wind-dominated regions when CTES is available provided waste heat is not utilized. This discrepancy can be attributed to the superior performance of hydrogen storage according to the techno-economic assumptions in [4], as it benefits from lower CPE, the absence of standing losses, and higher round-trip efficiency compared to PHPS.

PHPS is also selected in solar-dominated countries, particularly when waste heat is utilized. In such cases, PHPS can coexist with CTES, though its role becomes more prominent when paired with ITES systems. This configuration aligns with the prevalence of individual heating solutions, such as decentralized electric and gas boilers, common in these regions [33]. PHPS thus emerges as an optimal option for building-integrated applications producing both electricity and heat while storing waste heat in ITES. Large-scale buildings with substantial energy demands, such as hotels, sports centres, or industrial facilities that lack sufficient space for CTES, are well-positioned as early adopters, benefiting from the relatively large storage capacities and high energy density of PHPS units [15]. Alternatively, PHPS could be deployed at a large scale in solar-dominated countries as an electricity-only storage solution without waste heat recovery; however, its potential in this configuration is more limited (Table 2).

It is worthy emphasizing that PHPS systems that leverage waste heat emerges as a viable solution in solar-dominated countries when paired within the centralized configuration (i.e., operating as a large, centralized CHP plant). Although, district heating and cooling systems are currently not widely deployed in such regions in Europe, some researchers suggest its potential in Mediterranean countries [48]. District heating is also present in Middle Eastern European countries, which have greater solar resource than northern countries, and could thus be suitable for integrating PHPS. While district heating in other countries like the United States is more prominent in northern regions, it is also present in southern states [49]. In these contexts, PHPS could contribute to the energy system in the centralized configuration. By combining with CTES, PHPS can generate waste heat to supply heating demand while providing the system with a LDES. This functionality complements the short-duration operation of Li-ion batteries, creating a more resilient and balanced energy system.

## 6. Limitations

Although the results align with the literature and validate our findings, some limitations must be noted due to the assumptions of the study and the model used. For the sake of simplicity, electricity is mainly supplied by solar and wind generation even though other renewable sources are expected in the future energy system. It is assumed that OCGT, whose use is limited by the 5% of the $CO_2$ emissions constraints, represents other dispatchable power generators such as biomass, hydropower, etc. The same occurs with energy storage, only PHPS and Li-ion batteries are considered as electrical stores to balance wind and solar fluctuations. This study aims to understand the use of PHPS in the energy system, the most critical parameters that may hinder its prevalence, and the importance of the waste heat use for its feasibility. Thus, the assumed simplified model allows answering these questions qualitatively. Further studies identifying the energy storage characteristics that complement Li-ion batteries and PHS, and addressing current gaps of the energy system, have already been carried out by Gøtske et al. [36].

Space and time-resolution may also have an impact on the results. Each country simulation is represented by a single node; therefore, energy transmission from other countries is not considered. This limitation may impact on the PHPS' feasibility, potentially leading to an oversized optimal capacity. This effect could be more significant in wind-dominated scenarios where transmission lines are often used to manage variations in wind generation. The opposite effect might occur given that renewable energy exports are also not considered in the simulations, which is translated into larger energy curtailment. Additionally, bottlenecks within the national transmission lines are neglected, resulting into lower optimal energy storage capacities. Regarding time resolution, fast variations (<1 h) are not addressed as they fall out of the scope of this study. Consequently, fast-demand response technologies, such as flywheels or supercapacitors [8] are not analysed and do not compete with technologies under analysis. Furthermore, relying on a single typical year neglects the inter-annual variability of solar and wind generation, which could impact on the presented results [50].

Our model represents the heating and electricity coupling but ignores potential synergies with the industrial and transportation sectors. The industrial sector with high-temperature demand could benefit from the heat stored in PHPS.

## 7. Conclusions

This study has examined the role of PHPS systems in future renewable-dominated energy systems by embedding them into stylized national models for wind- and solar-dominated countries. The results demonstrate that PHPS can play an important role in providing long-duration flexibility, particularly in settings where CTES is not available. In such contexts, PHPS emerges as a cost-effective option for balancing low-frequency renewable fluctuations thanks to its low cost per energy capacity.

The analysis also highlights the decisive influence of waste heat utilization. When the low-temperature heat released during the power conversion process is recovered, PHPS becomes substantially more valuable, in some cases fully displacing lithium-ion batteries and lowering the system's overall dependence on short-duration storage. In solar-dominated regions, this coupling even allows PHPS to coexist with CTES, although its capacity remains much smaller in comparison. Conversely, in wind-dominated regions with established district heating systems, CTES tends to outcompete PHPS, limiting its role to scenarios where decentralized heating or waste heat recovery make it competitive.

Another notable finding is that PHPS can flexibly balance both synoptic-scale wind fluctuations and daily solar cycles, reflecting the advantage of decoupling its charging and discharging power capacities. However, its widespread deployment depends on achieving very low energy capacity costs—below 10 €/kWh—and maintaining reasonable heat-to-power conversion efficiencies. Cost reductions in lithium-ion batteries present an additional challenge, further narrowing the circumstances in which PHPS will be competitive.

Overall, the study suggests that PHPS is most promising in solar-dominated countries with decentralized heating systems or large-scale buildings where space for centralized storage is limited. In these contexts, PHPS can simultaneously deliver electricity and heat, reduce reliance on lithium-ion batteries, and contribute to system reliability.

## 8. Glossary

| Nomenclature | Meaning |
|---|---|
| AC | Alternate Current |
| av.h.l | Average hourly load |
| CAES | Compressed Air Energy Storage |
| CF | Capacity Factor |
| CHP | Combined Heat and Power |
| CPE | Cost per Energy |
| $CPP_{ch}$ | Cost per charge Power |
| $CPP_{dis}$ | Cost per discharge Power |
| CTES | Centralized Thermal Energy Storage |
| ETES | Electro Thermal Energy Storage |
| HV | High Voltage |
| ITES | Individual Thermal Energy Storage |
| LDES | Long-duration energy storage |
| LHTPV | Latent Heat Thermophotovoltaic |
| Li-ion | Lithium Ion |
| OCGT | Open Cycle Gas Turbine |
| PHPS | Power-to-Heat-to-Power |
| PHS | Pumped Hydro Storage |
| PV | Photovoltaic |
| SOC | State of Charge |
| TPV | Thermophotovoltaic |
| VRE | Variable Renewable Energy |

## 9. Acknowledgements

This work has been partially funded by the European Union's Horizon Europe research and Innovation Programme under grant agreement No 101057954 (THERMOBAT), and from the European Union's Horizon 2020 Research and Innovation Programme "SDGine for Healthy People and Cities" in UPM under the Marie Sklodowska-Curie agreement No. 945139. Views and opinions expressed are however those of the authors only and do not necessarily reflect those of the European Union or European Innovation Council. Neither the European Union nor the granting authority can be held responsible for them.

Figure 2 contains icons designed by Sumitsaengtong (PV), Freepik (wind turbine, buildings, thermometer and ITES), Assia Benkerroum (transmission tower), Smashicons (thermal power plant), IconBaandar (AC converter), Iconjam (battery), kosonicon (CTES), all of them retrieved from flaticon.com.

## 10. Author contributions

Conceptualization: A.D., M.V.; Data Curation: A.L.C, M.V, Funding Acquisition: A.D; I.A., C.d.C., Investigation: A.L.C., A.D., M.V; Methodology: A.L.C., M.V; Resources: A.D, M.V; Software: A.L.C, Visualization: A.L.C., M.V; writing – original draft: A.L.C., M.V; writing – review & editing: A.L.C., A.D, I.A., C.d.C., M.V.

## 11. Appendix A: Cost optimization model

The capacities of the components and their optimal dispatch are optimized to supply the energy demand in every hour (see Eq. (A.1)) while minimizing the total annualized system cost (see Eq. (A.2) below):

$$\sum_s g_{s,t} + \sum_s \alpha_{l,t} \cdot f_{l,t} = \sum_s d_t \leftrightarrow \lambda_t \quad (A.1)$$

Where the total energy demand in that bus (e.g., electricity), $d_t$, must be met at every timestep $t$ by means of the energy generated $g_{s,t}$ by each generator $s$ (including energy storage units) and energy from each link $l$, $f_{l,t}$, multiplied by the efficiency of that link at each timestep, $\alpha_{l,t}$. $\lambda_t$ is the Lagrange multiplier, indicating the marginal price of the energy carrier at each timestep.

$$\min_{G_s, E_s, g_{s,t}} \left[ \sum_s c_s \cdot G_s + \sum_s \hat{c}_s \cdot E_s + \sum_l c_l \cdot F_l + \sum_{s,t} o_{s,t} \cdot g_{s,t} \right] \quad (A.2)$$

Where $c_s$ is the annualized investment cost of each generator with $G_s$ power capacity, and $\hat{c}_s$ the annualized cost of the energy capacity $E_s$ for each energy storage technology $s$ in that node, $c_l$ is the annualized cost for each link $l$ with capacity $F_l$, and $o_{n,s,t}$ is the operational cost for each generator and storage technology dispatch, $g_{s,t}$, in every time-step $t$. Links include connections between different components and/or nodes (e.g., , heat pumps, AC-DC converters).

There are some constraints associated to the correct operation of the system. Power generation, $g_{s,t}$, cannot be greater than the generator power capacity, $G_s$ (see Eq. (A.3)), as well as power through the links, $f_{l,t}$, cannot surpass their maximum capacity, $F_l$, (see Eq. (A.4)). In addition, the state of charge (SOC) of the storage unit at each timestep, $e_{s,t}$, is determined by the stand-by losses ($loss$) as a percentage of the SOC in the previous timestep, $e_{s,t-1}$, the input power, $g_{s,t\ ch}$, with certain charging efficiency, $\eta_{ch}$; and the output power, $g_{s,t\ dis}$, with certain discharging efficiency, $\eta_{dis}$ (Eq. (A.5)). The SOC is limited to the maximum energy capacity, $E_s$, as well as the charging and discharging power is limited to their maximum capacities, $G_{s\ ch}$, $G_{s\ dis}$, respectively, and the energy balance within the store units (see Eq. (A.3)).

$$0 \leq g_{s,t} \leq G_s \tag{A.3}$$

The maximum power, $g_{s,t}$, through each link $l$ is limited to their maximum power capacity $G_s$.

$$0 \leq f_{l,t} \leq F_l \tag{A.4}$$

Equivalently, the maximum power, $f_{l,t}$, through each link $l$ is limited to their maximum power capacity $F_l$.

$$e_{s,t} = e_{s,t-1} - loss \cdot e_{s,t-1} + \eta_{ch} \cdot g_{s,t\ ch} \cdot \Delta t - \eta_{dis}^{-1} \cdot g_{s,t\ dis} \cdot \Delta t, \qquad 0 \leq e_{s,t} \leq E_s \tag{A.5}$$

Other constraints are related to the configuration we aim to simulate, as the limit in the $CO_2$ emissions, defined as per Eq. (A.6), and the maximum ITES energy capacity, Eq. (A.7).

$$\sum_s \varepsilon_s \cdot \frac{g_{s,t}}{\eta_{s,t}} \leq CAP_{CO_2} \leftrightarrow \mu_{CO_2} \tag{A.6}$$

Where the total $CO_2$ emissions of the country are calculated from specific emissions in $CO_2$-tonne-per-MWh$_{th}$ of the generator $s$, $\varepsilon_s$; the energy generated in each timestep $t$ for that generator and the efficiency of the generator $s$, $\eta_{s,t}$. These are limited to the maximum $CO_2$ emissions $CAP_{CO2}$. The shadow price for the $CO_2$, $\mu_{CO_2}$, is obtained from this constraint, representing the price of $CO_2$ that should be added to obtain a system with the required emissions.

$$E_{ITES} \leq E_{ITES\ max} \tag{A.7}$$

Where $E_{ITES}$ is the resulting optimum energy capacity of the ITES, limited to the maximum set capacity, $E_{ITES-max}$.

Only the electricity and heating sectors are considered in this study and are represented by the AC Electrical and Heating buses respectively. The AC Electrical bus gathers both the High and Low Voltage systems. The electrical demand time series has been retrieved from European Network Transmission System Operators for Electricity (ENTSO-e) through the compilation by Open Power System Data [40] for the year 2015. The electricity used to cover electrified domestic heating demand is subtracted as it needs to be included in the heating demand. The definitive yearly electricity demand for Spain and Denmark are 248 TWh and 33 TWh respectively. Solar PV, onshore wind installations and OCGT plants supply the energy demand for that node. Hourly data series for Capacity Factor of solar PV and wind have been obtained from [38], [39]. $CO_2$ emissions are capped at 5 % of the emissions in 1990 associated to electricity and heating demand for each country. This results into annual $CO_2$ emissions of 7 Mt for Spain and 2 Mt for Denmark.

Heating bus encompasses both district heating systems and individual installations. Space heating demand is estimated through the Heating Degree Hours (HDH) method, which relies on hourly ambient temperature data. This data, sourced from [51], provides information for each cell division within each country's geographical space in the database. After the HDH calculation, a weighting based on the population in each region in the country is applied and aggregated to estimate total space heating demand of that country. Domestic hot water demand time series is included in the same database. The final annual heating demand for Spain and Denmark are 156 TWh and 54 TWh respectively.

## 12.     Appendix B: Results for the solar-dominated location (Spain)

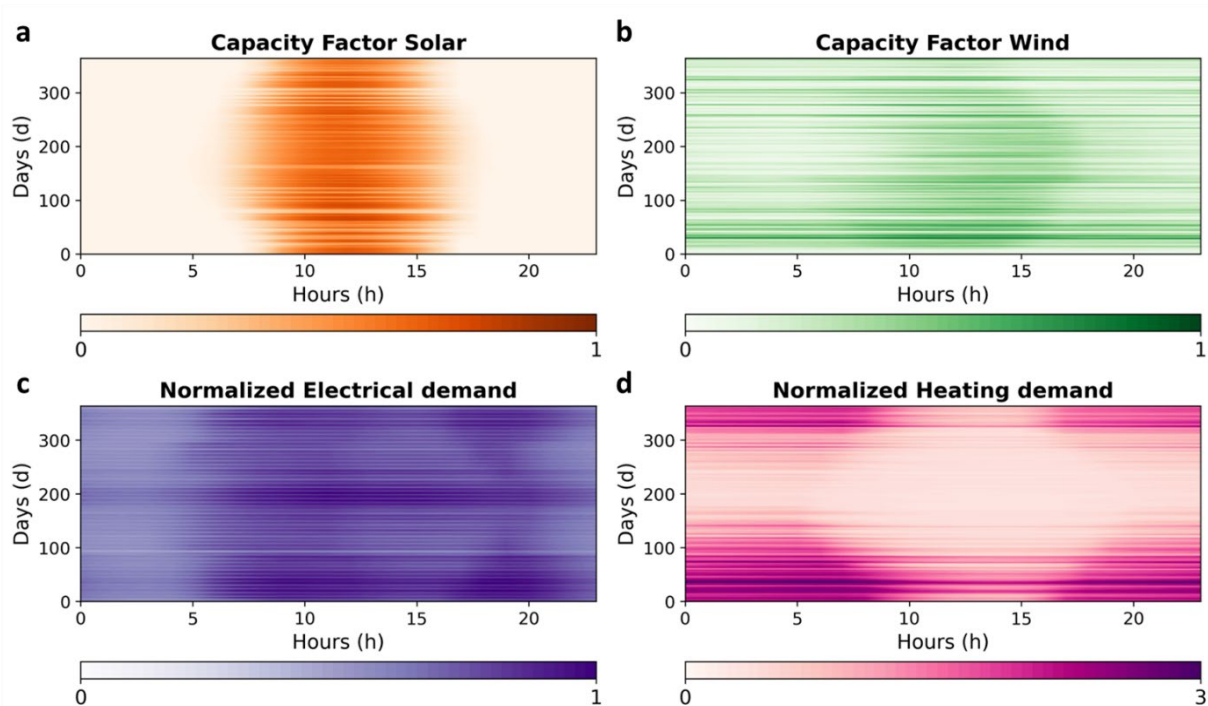

Figure B. 1 Panels a-d plot hourly data during the hours of the day (x-axis) and days of the year (y-axis) for the solar-dominated location, Spain. Panel a and b represent the solar and wind

capacity factors respectively. Panels c and d contain the normalized electrical and heating demand relative to their respective average demand.

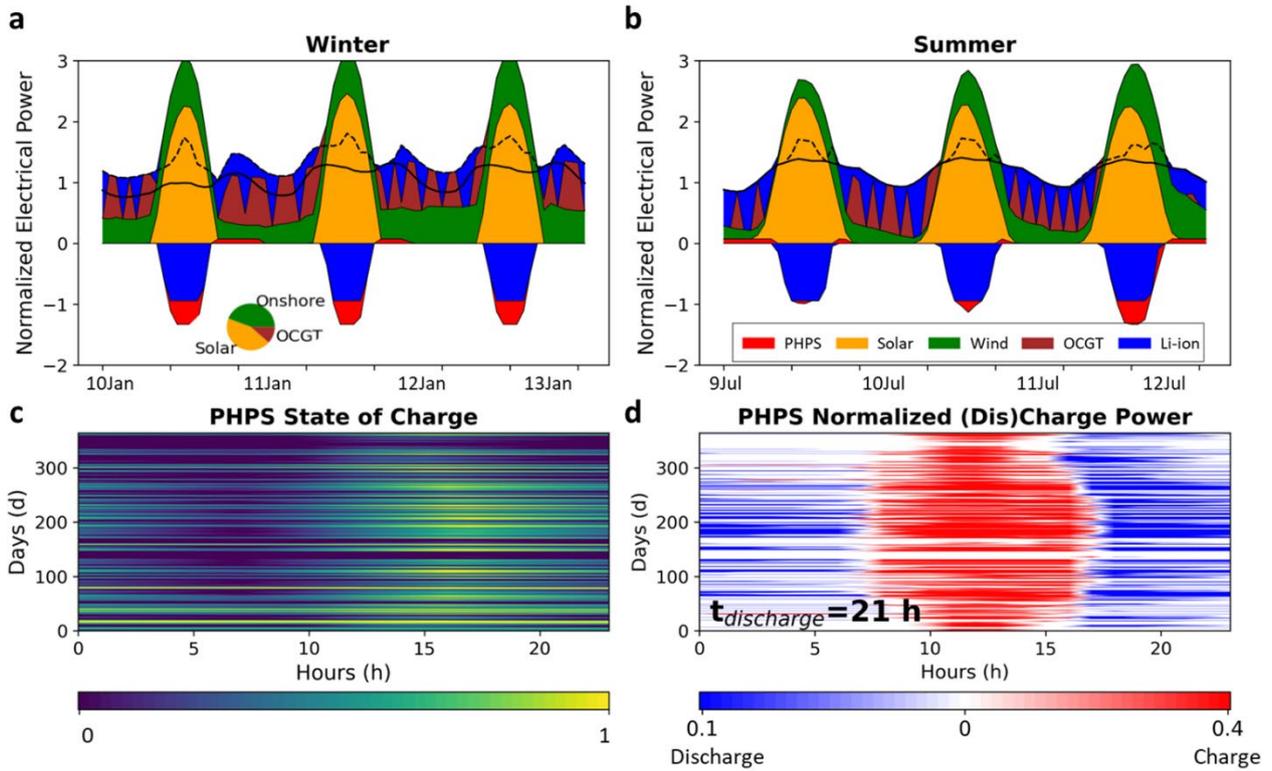

Figure B. 2. Energy balance and storage operation for the solar-dominated location, not considering waste heat in the decentralized configuration. All the panels in this figure show values normalized to the average electrical demand for Spain (28 GW). Panels a and b depict the normalized electrical generation (positive values) of wind (green), solar PV (yellow), OCGT (brown), PHPS (red) and Li-ion batteries (blue) for three days in winter (panel a) and summer (panel b). Negative values represent the power used to charge the PHPS (red) and Li-ion batteries (blue). The normalized exogenous electricity demand is plotted in a thick black line. The normalized total electricity, represented with dashed black line, includes electricity demand and heat pump and heat resistance electrical consumption. The total electricity demand (including electricity consumed by the heat pump and resistive heater) is plotted in a dotted black line. The annual electricity mix for this scenario is plotted in a pie-chart, inserted in panel b. Panel c shows the normalized state of charge of the PHPS throughout the day (x-axis) and year (y-axis), being yellow when completely charged and dark-blue when discharged ($E_{cap-PHPS}$=104 GWh). Panel d shows the normalized charging (red) and discharging (blue) power of the PHPS throughout the day (x-axis) and year (y-axis). Its discharge time at maximum discharging power capacity, $E_{capac-PHPS}$ (kWh)·$\eta_{disch-PHPS}$/$P_{el-disch-PHPS}$ (kW), is 20 hours.

Table B. 1 Optimal Li-ion battery, CTES and ITES energy capacities represented as the number of av.h.l (h/y) that the storages could supply the average power demand in the system. This parameter is used as an indication of the relevance of the PHPS storage.

| Location | Wind-dominated | | | | Solar-dominated | | | |
|---|---|---|---|---|---|---|---|---|
| | No waste heat | | Waste heat | | No waste heat | | Waste heat | |
| Configuration / Energy storage | Li-ion | CTES/ITES | Li-ion | CTES/ITES | Li-ion | CTES/ITES | Li-ion | CTES/ITES |
| Base | 4 | 0 | 0 | 0 | 8 | 0 | 5 | 0 |

| Decentralized | 1 | 13 | 0 | 5 | 6 | 6 | 4 | 3 |
|---|---|---|---|---|---|---|---|---|
| Centralized | 0 | 588 | 0 | 589 | 5 | 180 | 4 | 139 |

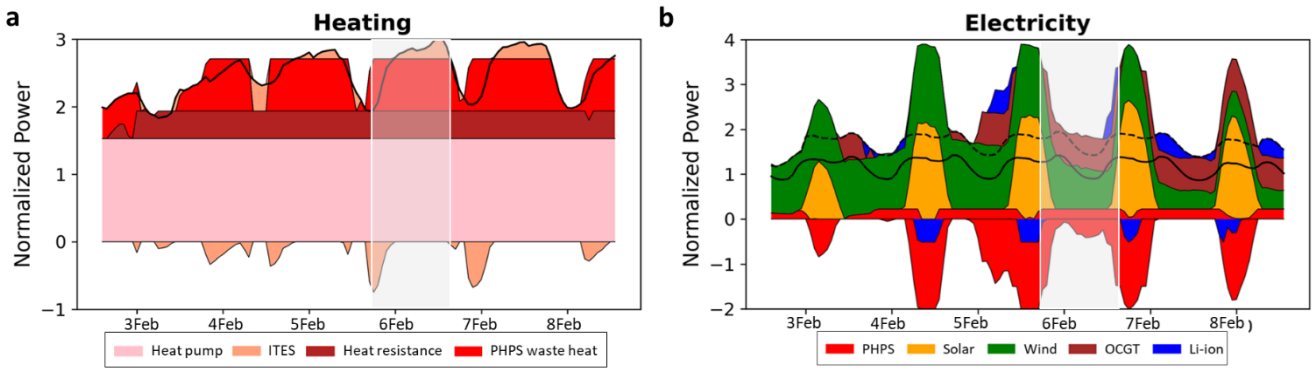

Figure B. 3. Results for the solar-dominated location, considering use of waste heat in the decentralized configuration normalized to the yearly average heating and electrical demand of Spain ($P_{th-avg-ye}$=18 GW, $P_{el-avg-ye}$=28 GW). Panel a depicts the heat supplied by heat pump (pink), heat resistance (brown), PHPS waste heat (red) and ITES (orange positive region), and the charging of ITES (negative orange region). Panel b depicts the electrical generation (positive values) of wind (green), solar (yellow), OCGT (brown), and discharge of PHPS (red) and Li-ion (blue) for 5 days in winter. Negative values represent the power used to charge the PHPS (red) and Li-ion batteries (blue). The instant heating and electricity demand are plotted in a thick black line in their respective panels. The normalized total electricity, represented with dashed black line, includes electricity demand and heat pump and heat resistance electrical consumption. The grey area marked in both panels exemplifies a timeslot when charging and discharging of PHPS occurs at the same time (panel b).

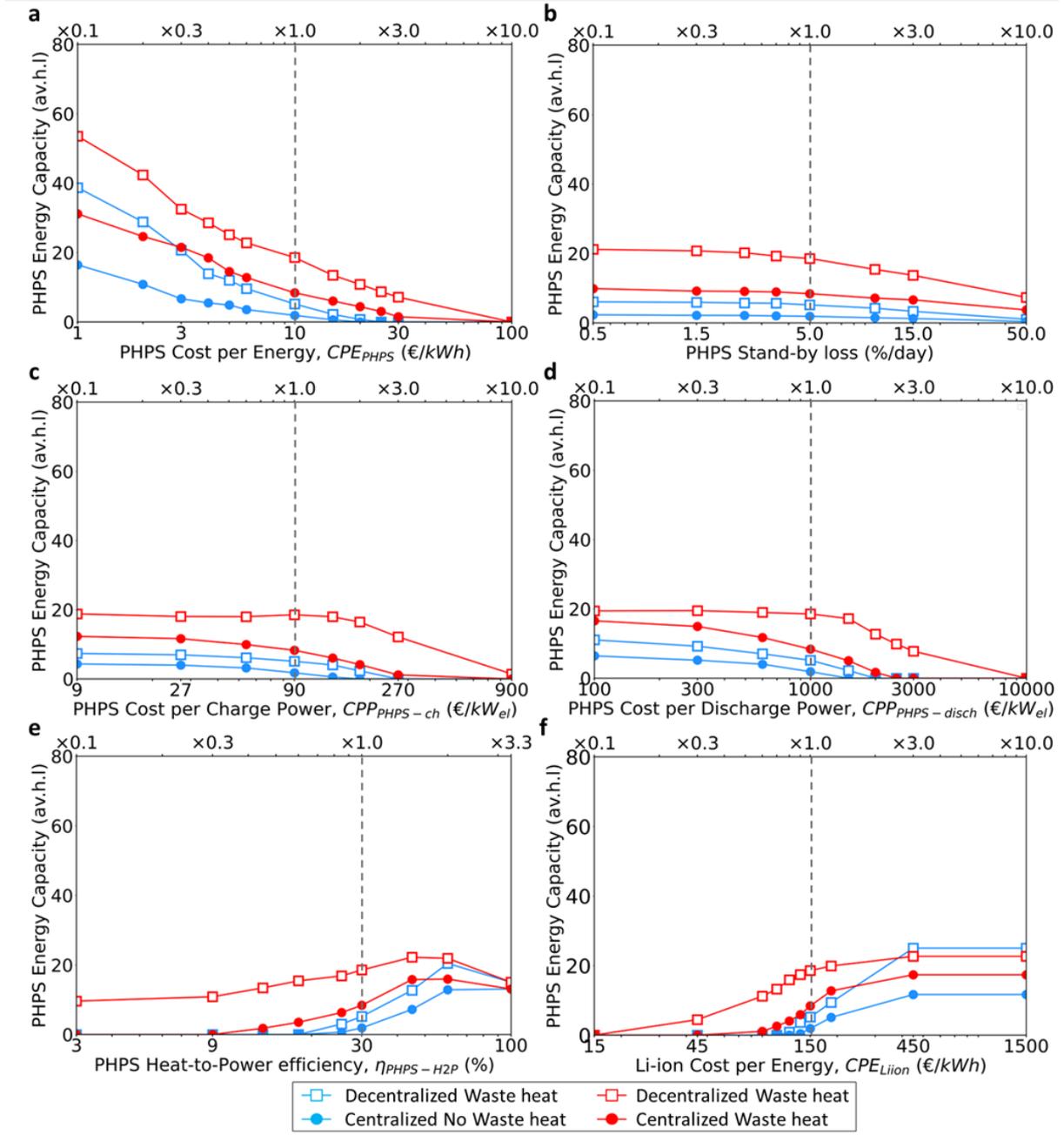

Figure B. 4. Optimal PHPS energy capacity for solar-dominated system, expressed in av.h.l vs different PHPS parameter assumption: (a) Cost per Energy (CPE, €/kWh), (b) stand-by losses (%/day), Cost per (c) Discharge and (d) Charge electrical Power ($CPP_{dis}$, $CPP_{ch}$, €/kW), (e) Heat-to-Power conversion efficiency (H2P, %), and (f) Li-ion cost variation. The reference techno-economic assumptions in Table 1 are depicted by the dashed grey line.